\documentclass{INTERSPEECH2023}

\usepackage{caption}
\usepackage{subcaption}
\usepackage{multirow}
\usepackage{multicol}
\usepackage{algorithm}
\usepackage{algpseudocode}

\newcommand{\ed}[2]{\left\Vert #1-#2 \right\Vert_2}
\newcommand{\inparanth}[1]{\left(#1\right)}
\newcommand{\R}{\mathbb{R}}

\interspeechcameraready

\title{Matching Latent Encoding for Audio-Text based Keyword Spotting}
\name{ Kumari Nishu, Minsik Cho, Devang Naik}
\address{
Apple}
\email{knishu@apple.com, minsik@apple.com, naik.d@apple.com}

\begin{document}

\maketitle
 
\begin{abstract}
Using audio and text embeddings jointly for Keyword Spotting (KWS) has shown high-quality results, but the key challenge of how to semantically align two embeddings for multi-word keywords of  different sequence lengths remains largely unsolved.
In this paper, we propose an audio-text-based end-to-end model architecture for flexible keyword spotting (KWS), which   builds upon   learned audio and text embeddings. 
Our architecture uses a novel dynamic programming-based   algorithm, Dynamic Sequence Partitioning (DSP), to optimally partition the audio sequence into the same length as the word-based text sequence using the monotonic alignment of spoken content.
Our proposed model consists of  an encoder block to get audio and text embeddings, a projector block to project individual embeddings to a common latent space, and an audio-text aligner containing a novel DSP algorithm, which aligns  the audio and text embeddings to determine if the spoken content is the same as the text.
Experimental results show that
our DSP is more effective than other partitioning schemes, and 
the proposed architecture outperformed the state-of-the-art results on the public dataset in terms of Area Under the ROC Curve
(AUC) and Equal-Error-Rate (EER)  by   14.4 \% and 28.9\%, respectively.


\end{abstract}


\begin{figure*}[t!]
  \centering
  \includegraphics[width=\textwidth]{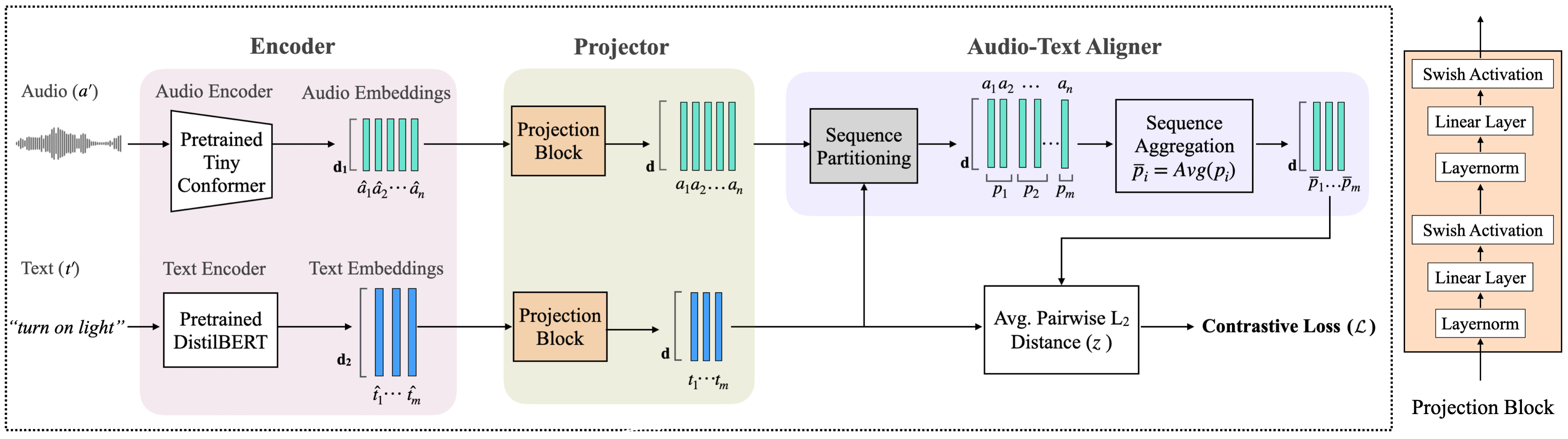}
  \caption{Overall architecture of the proposed end-to-end KWS model i.e EMKWS (Encoding Matching based KeyWord Spotter). It has three modules- encoder, projector, audio-text aligner. Input to the model- (audio and text) pair, output- a binary label indicating if the audio content is same as the text.}
  \label{fig:model}
\end{figure*}

\section{Introduction}

Voice assistants on smart devices, like smart phones and small speakers, have become a ubiquitous part of our daily life
by providing convenience in controlling personal/home devices while multi-tasking (e.g., driving).
Thus,   natural human-like interactions with such voice assistants are in greater demand than ever. Voice assistants commonly rely on Keyword Spotting (KWS) techniques to detect whether a custom trigger phrase  and voice command  was uttered in an input speech signal.

There are two broader categories of KWS tasks: fixed vocabulary KWS, where models work only for a fixed list of keywords \cite{Chen2014SmallfootprintKS,Tang2017DeepRL}, and flexible KWS, where models are expected to work for any arbitrary keyword enrolled by users \cite{qbe1,Generalized_qbe2, shin22_interspeech}.
Flexible KWS is a highly challenging task as it involves detecting arbitrary keywords, which may not have been observed during model training. In flexible KWS, a user can specify their custom keywords either in the form of enrolled audio samples (also known as query-by-example, QbE KWS \cite{qbe1,Generalized_qbe2})  or as the text.
In this work, we  focus on the latter where a custom keyword is provided as text, and a sequence of words.




Prior works in flexible KWS built on embedding-based neural network modeling have shown promising results \cite{embedding1,embedding3}, but suffer from being inapplicable for heterogeneous modalities (like audio and text)~\cite{dtw,dtw_using,dtw_using3}, being ineffective on multi-word keyword detection \cite{Generalized_qbe2,He2016MultiviewRN,Word2Vec}, or failing to 
capture word-level semantics~\cite{openVocabularyKS}. For more details, please refer to Section~\ref{related_work}.



To address such challenges, we propose a novel embedding-based neural network architecture for  flexible KWS, which consists of encoders, projectors, and an audio-text aligner. 
Unlike   existing projection approaches in \cite{Generalized_qbe2,He2016MultiviewRN,Word2Vec}, our projection does not alter the sequence length but focuses on transformation of embeddings into a common space. Then, our audio-text aligner, powered by a novel Dynamic Sequence Partitioning (DSP) algorithm, finds   the best audio sequence partitioning for high-quality audio-to-text mapping. Finally, all of the audio frames in each partition are aggregated for distance computation from the paired word in the text.  Such a distance can be used to determine whether the target keyword (in text) also exists in the audio sequence.
Our major contributions are  as follows:
\begin{itemize}
    \item We propose a novel Dynamic Sequence Partitioning (DSP) algorithm to compare two sequences of different lengths where \textbf{a)} contents are monotonically aligned, \textbf{b)} the indices from the shorter sequence have one-to-many relationship with the longer one, but \textbf{c)} the indices from the longer sequence have one-to-one relationship with the shorter one.
    \item  We present a flexible embedding matching setup to map an acoustic embedding sequence  to its corresponding textual word embedding sequence, even from two independently pre-trained encoders, such that the audio sequences for the same text are embedded similarly.
    \item We formulate a discriminative setting to train an end-to-end KWS model for both single and multi-word keywords. Our model consists of encoders, projectors and a novel DSP-based audio-text aligner.

\end{itemize}

\section{Related work}
\label{related_work}

One approach for flexible KWS is to use a well-known dynamic time warping (DTW) algorithm \cite{dtw} to measure the distance between two embedding sequences of different lengths for the KWS task \cite{dtw_using,dtw_using3}.  DTW calculates the best match between two given sequences, where the indices of one sequence have a one-to-many relationship with the indices of the other sequence, which works well when we compare homogeneous modalities (like audio vs. audio) since an index from both ends represents the same entity (e.g., $25ms$ audio frame per index). However, DTW becomes less applicable or inaccurate when we compare heterogeneous modalities, such as audio (where each index represents one audio frame) and text (where each index represents one word token), which requires one index of a word to be mapped to multiple indices of audio frames but not vice versa.
To overcome such limitations, \cite{Generalized_qbe2,He2016MultiviewRN,Word2Vec} 
project both audio and text sequences into a fixed sequence length which  makes the distance computation between two sequences simpler at the cost of losing critical semantic context  during projection (i.e., projection does dimensionality reduction). Such projection-based schemes have been shown to learn word discrimination better than the DTW-based schemes but tend to deliver poor performance on multi-word keyword detection. Recently, \cite{openVocabularyKS}   explored an audio-text based flexible KWS in a   different way, where
audio and text encoders generate flexible length sequences of embedding vectors per phoneme, which is fed into a GRU network to
generate a single embedding vector for both the audio and text modalities. However, such phoneme-based embeddings do not effectively learn to discriminate words as they fail to capture word-level semantics. 

The recent work in \cite{shin22_interspeech} measures the audio and text correspondence for   flexible KWS task and have reported the state-of-the art results compared to  CTC \cite{Graves2006ConnectionistTC} and attention \cite{Vaswani2017AttentionIA} based methods. However, the text encoder of \cite{shin22_interspeech} does not capture the semantics of the textual keyword as it is based on the grapheme and phoneme sequence, hence 
\cite{shin22_interspeech} is weak against similar sounding.
 Moreover, \cite{shin22_interspeech} learns explicit patterns between text and audio embeddings with monotonic matching loss,  which compromises the alignment flexibility and hence it might not end up with fully optimized alignment.

\section{Proposed Method}
In this section, we describe our scheme, Encoding Matching based Keyword Spotter (EMKWS) shown in Fig. \ref{fig:model}.
EMKWS consists of three modules - an encoder, a projector, and an audio-text aligner, which drive  our contrastive loss function.
EMKWS model takes an audio and text pair as input and measures the similarity between the spoken content of the audio and the text to determine if they match. Let us denote an input example as $\inparanth{a',t',l'}$, where audio $a'=\inparanth{a'_1, a'_2, \dots, a'_{n'}}$ is a sequence of audio frames, text $t'=\inparanth{t'_1, t'_2, \dots, t'_{m'}}$ is a sequence of text tokens, and $l'$ is the ground truth binary label with $l'=1$ for a positive input pair and $0$ otherwise.
We will loosely refer to text as a sequence of words for better clarity.  The rest of this section describes each module in more detail.

\subsection{Encoder}
The encoder module consists of an audio encoder and a text encoder as shown in Fig. \ref{fig:model}.
 We use the Conformer architecture \cite{conformer} as the audio encoder
 to generate the embedding for a given audio input signal. This choice was inspired by the fact that the Conformer combines the power of self-attention \cite{Vaswani2017AttentionIA} to learn global interactions and convolutions to capture local correlations efficiently. This is particularly useful for speech data as speech contains sequential information along with the global context in which the speech is being delivered. We employ a tiny version of the Conformer architecture described in Section~\ref{results}. We first train this audio encoder for Automatic Speech Recognition (ASR) to produce a transcription for the given audio, we remove the last linear layer, and then take the remaining architecture as a frozen audio encoder. Given $a'$ as the input, let us denote the output from this audio encoder as $\hat a=\inparanth{\hat a_1, \hat a_2, \dots, \hat a_n}$, where $\hat a_i\in\R^{d_1}, \forall\, i=1,2,\dots, n$ and $n < n'$ as the Conformer module has a subsampling layer to reduce the encoded sequence length.

For the text sequence, we use a transformer-based model trained for natural language understanding, DistilBERT \cite{DistilBERTAD}, to obtain a semantically focused text embedding for the input text. Given $t'$ as the input text, let us denote the output from this text encoder as $\hat t=\inparanth{\hat t_1, \hat t_2, \dots, \hat t_m}$, where $\hat t_k\in\R^{d_2}, \forall\, k=1,2,\dots, m$ and $m \geq m'$ as the tokenizer of DistilBERT may produce multiple sub-words as tokens for a single word.

\subsection{Projector}
The incoming audio and text embeddings from the encoder module are passed through the respective projection block in the
projector module, as in Fig. 1. So far, the two modalities have been encoded in their separate embedding spaces. The role of the projector module is to project both embeddings to a common embedding space of dimension $d$. The projection block consists of two layer norms and two linear layers,  as shown on the right side of Fig. \ref{fig:model}. Given the audio and text embedding inputs $\hat a$ and $\hat t$, we denote the projected audio embedding as $a=\inparanth{a_1, a_2, \dots, a_n}$, where $a_i \in \R^d, \forall\, i=1,2,\dots,n$ and the projected text embedding as $t=\inparanth{t_1, t_2, \dots, t_m}$, where $t_k \in \R^d, \forall\, k=1,2,\dots, m$.

\subsection{Audio-Text Aligner}
Our audio-text aligner is composed of two parts, sequence partitioning and sequence aggregation.
The audio and text embeddings generated from the projector module have the same embedding dimension $d$ at the different sequence lengths $n$ and $m$ in Fig. \ref{fig:model} (and, $n>m$ in most cases), and the aligner transforms the audio embedding into the same dimension as the text embedding (i.e., $\R^{m \times d}$) to compare them. Even though their sequence lengths are different, the content alignment between the two is monotonic in nature.
Our {Dynamic Sequence Partitioning (DSP) described in Algorithm~\ref{alg:dppartitioning} leverages such monotonic characteristic to perform sequence partitioning of the audio sequence $a=\inparanth{a_1, a_2, \dots, a_n}$ into $m$ chunks. Then, each partition is aggregated for loss generation.

Let us denote a partition of the sequence $a$ into $m$ contiguous sub-sequences as $p=\inparanth{p_1, p_2, \dots, p_m}$, as shown in the Fig. \ref{fig:model}. This partition can be represented using the $(m+1)$ indices $i_0, i_1, \dots, i_m$ (where $i_0=1$, $i_m=n+1$) such that $p_1 = a_{i_0:i_1}$, $p_2 = a_{i_1:i_2}$, and similarly $p_m = a_{i_{(m-1)}:i_{m}}$, where we use the notation $a_{i:j}$ to denote the sub-sequence $\inparanth{a_i, a_{(i+1)}, \dots, a_{(j-1)}}$. Then, during the next sequence aggregation step, we aggregate each sub-sequence to create output as $\inparanth{\overline{p_1}, \overline{p_2}, \dots, \overline{p_m}}$ where over-line $\overline{\inparanth{.}}$ denotes the mean of a sequence of vectors. Note that the output is a sequence containing $m$ vectors of $d$ dimension each, and thus can be compared directly with the text embedding $t$. 
Finally, we measure the distance between these two sequences as the average pairwise $L_2$ distance, i.e. $\frac{1}{m}\sum_{k=1}^m \ed{\overline{p_k}}{t_k}$.

Since the model training is guided by such $L_2$ distance, our method captures the best possible partitioning solution by directly minimizing the $L_2$ distance using 
\textbf{Dynamic Sequence Partitioning (DSP) algorithm} in Algorithm~\ref{alg:dppartitioning}. DSP 
finds the optimal partition such that the aggregated audio embedding and the text embedding have minimum average pairwise $L_2$ distance $z$ calculated as in the following formulation:
\begin{equation} \label{eq:optimalpartition}
    z(a,t) = \min_{p \in Partition(a,m)} \inparanth{\frac{1}{m}\sum_{k=1}^{m}\ed{\overline{p_k}}{t_k}}
\end{equation}
where $Partition(a,m)$ denotes the set of all possible partitions of sequence $a$ into $m$ chunks.

Our optimal mapping from the audio to the text embeddings will offer the best possible view of the mapping throughout the training, and our experimental results in Fig.~\ref{fig:partition} suggest that our approach is effective, especially for the hard samples.
The computational complexity of DSP is $\mathcal{O}(mn^2)$ but since $m$ is usually very small ($m\leq 4$ in our case), the complexity becomes $\mathcal{O}(n^2)$. Similarly, the space complexity is $\mathcal{O}(n)$.

\begin{algorithm}[t]
\caption{Dynamic Sequence Partitioning (DSP) Algorithm}
\label{alg:dppartitioning}
\begin{algorithmic}
    \State \textbf{Input:}
        \begin{itemize}
            \item Audio Embedding, $a=\inparanth{a_1, a_2, \dots, a_n}$
            \item Text Embedding, $t=\inparanth{t_1, t_2, \dots, t_m}$
        \end{itemize}
    \State
    \State \textbf{Initialization:}
    \State \textbf{1. } Let $D \in \mathbb{R}^{m \times n}$, $D[k][i]=0$, $\forall\, 1\leq k \leq m, 1\leq i \leq n$
    \State $D[k][i]$ will store the optimal loss to divide the array $a_{i:(n+1)}$ in $k$ chunks to match with the last $k$ text tokens
    \State \textbf{2. } Set $D[1][i] = \ed{\overline{a_{i:(n+1)}}}{t_{m}}$, $\forall\, 1 \leq i \leq n$
    \State
    \State \textbf{Procedure:}
    \For {$k=2,3,\dots,m$}
        \For {$i=1, 2, \dots, n-k+1$}   
            \State Set $\delta_j=\ed{\overline{a_{i:j}}}{t_{(m-k+1)}}$, $\forall\, i< j \leq (n-k+2)$
            \\
            \State $\displaystyle{D[k][i] = \min_{i< j \leq (n-k+2)}  \inparanth{\delta_j + D[k - 1][j]}}$
        \EndFor
    \EndFor
    
    \State \textbf{Output: } $D[m][1] / m$, optimal loss to divide the sequence $a$ in $m$ chunks to best align with the $m$ tokens of sequence $t$
\end{algorithmic}
\end{algorithm}

\subsection{Contrastive Loss}
    We use   contrastive loss \cite{contrastiveloss} as our training objective. The contrastive loss takes the average pairwise $L_2$ distance, $z$, from Equation \ref{eq:optimalpartition} as input, which minimizes the distance for positive examples and maximizes the distance for negative examples. The loss $\mathcal{L}$ for an example $\inparanth{a',t',l'}$ is calculated as:
    \begin{equation}\label{eq:loss}
        \mathcal{L}=l'\cdot \max{\inparanth{z-m_{pos}, 0}} + (1-l') \cdot\max{\inparanth{m_{neg}-z, 0}}
    \end{equation}
    where $l'$ is the ground truth label which is $1$ for a positive example and $0$ for negative example, $z$ is the computed distance, $m_{pos}$ is the positive margin over which positive examples will contribute to the loss, and similarly $m_{neg}$ is the negative margin under which negative examples will contribute to the loss.

\section{Experimental Results}
\label{results}
We implemented our method in PyTorch, and experimented on a x86 Linux machines with  NVIDIA V100 GPU.

\subsection{Experiment Setup}
    \textbf{Dataset:} We trained and evaluated the proposed method using LibriSpeech \cite{librispeech} and Libriphrase datasets \cite{shin22_interspeech}. The audio encoder in Fig. \ref{fig:model} is trained with \textit{train-clean-100} and \textit{train-clean-360} from the LibriSpeech dataset using character based tokenization.
    Following \cite{shin22_interspeech}, we generated a phrase unit based Libriphrase dataset from LibriSpeech to train and evaluate the EMKWS model. The training set of Libriphrase was generated from \textit{train-clean-100} and \textit{train-clean-360}, and the evaluation set was generated from \textit{train-others-500}. The Libriphrase dataset consists of $1$, $2$, $3$, and $4$ word-phrases. Each example is denoted by $3$ entities: (\textit{audio, text, target}) where the target value is $1$ if the audio and text both represent the same content, otherwise $0$. There are two types of negative examples, easy and hard, based on the Levenshtein distances \cite{Levenshtein1965BinaryCC} between the audio content and the paired text. 
    To make comparison with \cite{shin22_interspeech}, we built the same evaluation dataset, Libriphrase Easy (LE) and Libriphrase Hard (LH), consisting of $4391$, $2605$, $467$, and $56$ episodes of each word length respectively. Each episode has $3$ positive and $3$ negative audio-text pairs. The audio features are extracted using $80$-channel filterbanks from a $25$ms window with a stride of $10$ms.

    \hspace{-0.2in }\textbf{Pre-trained Models:}  We used the Conformer architecture \cite{conformer} to build the audio encoder in Fig. \ref{fig:model}. The Conformer hyper-parameters are \textit{\{6 encoder layers, encoder dimension of 144, convolution kernel of size 3, and 4 attention heads\}}. Lastly, we added a {Linear layer} with output dimension of $29$ and trained the resulting model with CTC loss \cite{Graves2006ConnectionistTC}. The model was trained for $300$ epochs with Adam optimizer \cite{Kingma2014AdamAM} and transformer learning rate schedule \cite{Vaswani2017AttentionIA} with $5$k warm-up steps. After the training, we removed the {Linear layer} and froze the parameters of the audio encoder to use in EMKWS. For the text encoder, we use DistilBERT \cite{DistilBERTAD}, a smaller version of the natural language understanding model BERT \cite{bert}. We took the pretrained DistilBERT model from the Hugging Face library \cite{hugginface-transformers}. 
    
    \subsection{Training and Inference}
    
    We train the end-to-end EMKWS model with the frozen audio and text encoders. The output of encoder module gives audio and text embeddings of dimensions $d_1=144$ and $d_2=768$ respectively (see Fig.~\ref{fig:model}). The projector module projects them to a common embedding space of dimension $d=144$.
    For the contrastive loss in Eq. \ref{eq:loss}, we experimented with a combination of positive and negative margin hyper-parameters, with $m_{pos}\in\{0, 0.1, 0.2, 0.25, 0.4, 1.0\}$ and $m_{neg}\in\{ 2, 2.5, 4, 7, 10, 15\}$, and the best  was $m_{pos}=0.2$ and $m_{neg}=7$. EMKWS is trained for $165$ epochs with Adam optimizer \cite{Kingma2014AdamAM} and transformer learning rate schedule \cite{Vaswani2017AttentionIA} with $2$k warm-up steps.  
    EMKWS model uses the pre-computed embeddings of the enrolled keywords instead of keeping the text encoder during inference, which leads to $3.7$M parameters in total.

     \begingroup
    \renewcommand{\arraystretch}{1.25}
        \begin{table}[t]          
          \centering
          \begin{tabular}{l|l|l|l|l}
            \toprule
            \multirow{2}{*}{Method} & \multicolumn{2}{l|}{\textbf{AUC Score (\%)}} & \multicolumn{2}{l}{\textbf{EER (\%)}} \\
            & LE & LH &LE & LH\\
            \hline
            \textbf{Baseline} \cite{shin22_interspeech} & 96.7 & 73.58 & 8.42 & 32.9\\
            \hline
            Proposed (run 1) & 97.83 & 84.21 & 7.26 & 23.37\\
            Proposed (run 2) & 97.84 & 84.18 & 7.34 & 23.29\\
            Proposed (run 3) & 97.81 & 84.24 & 7.48 & 23.42\\
            \hline
            \textbf{Proposed (avg)} & \textbf{97.83} & \textbf{84.21} & \textbf{7.36} & \textbf{23.36}\\
            {\textbf{Improvement (\%)}} & \textbf{1.2} & \textbf{14.4} & \textbf{12.6} & \textbf{28.9}\\
            \bottomrule
          \end{tabular}
          \vspace{0.1 in}
          \caption{Experimental results of the proposed EMKWS model on Libriphrase Easy (LE) and Libriphrase Hard (LH) datasets with comparison to the baseline model.}
          \label{tab:result}
        \end{table}
    \endgroup
    
\subsection{Results and Discussion}
    We compared the proposed EMKWS model with the state-of-the-art results from baseline~\cite{shin22_interspeech}  in Table \ref{tab:result}, which shows that our method
    outperforms the baseline in terms of Area Under the ROC Curve (AUC) and Equal-Error-Rate (EER) metric on Libriphrase Easy (LE) and Libriphrase Hard (LH) datasets.  On the LE dataset, we see an improvement of $1.2$\% in AUC score and $12.6$\% in EER metric. The proposed model showed improvement on the LH dataset: AUC score improved by $14.4$\% and EER metric improved by $28.9$\%.

    To understand the benefit of our dynamic programming scheme, we experimented with three different sequence partitioning algorithms: random, equal, and our DSP, and  the experiment results are shown in Fig.~\ref{fig:partition}. Although there are limited multi-word examples in the evaluation datasets, we see the DSP algorithm consistently superior to the other  schemes.
    
    We also examined how each word embedding in the text sequence is aligned with the embeddings of their corresponding chunks of audio frames, and visualized the correlation between text-to-audio and audio-to-audio embeddings resulted from the projector module  in  Fig. \ref{fig:aligner}.
    We see three distinct blocks of strong monotonic correlation along the audio embedding axis, and each block is aligned with each word in Fig. \ref{fig:aligner} (a). 
    At the same time, Fig. \ref{fig:aligner} (b) shows that embeddings of audio frames for each word are fully self-correlated. 
    We conjecture that such intuitive patterns between the projected embeddings are due to the DSP algorithm. For the examples of  Fig. \ref{fig:aligner} (a) and (b), the input audio length was $550$ms with the words \textit{``the"}, \textit{``old"} and \textit{``man"} spoken for $90$ms, $230$ms and $230$ms, respectively.
    Thus, the relative portion of each word is approximately $\inparanth{0.16:0.42:0.42}$ which matches the partition sizes $\inparanth{2,5,5}$ from the DPS algorithm as highlighted with a dashed vertical line in Fig. \ref{fig:aligner} (a) and (b).
    We can make similar observations  for the four word positive example \textit{``of the United States"} in Fig. \ref{fig:aligner} (c) and (d).

   

    

    \begin{figure}[t!]
      \centering
      \includegraphics[width=0.95\linewidth]{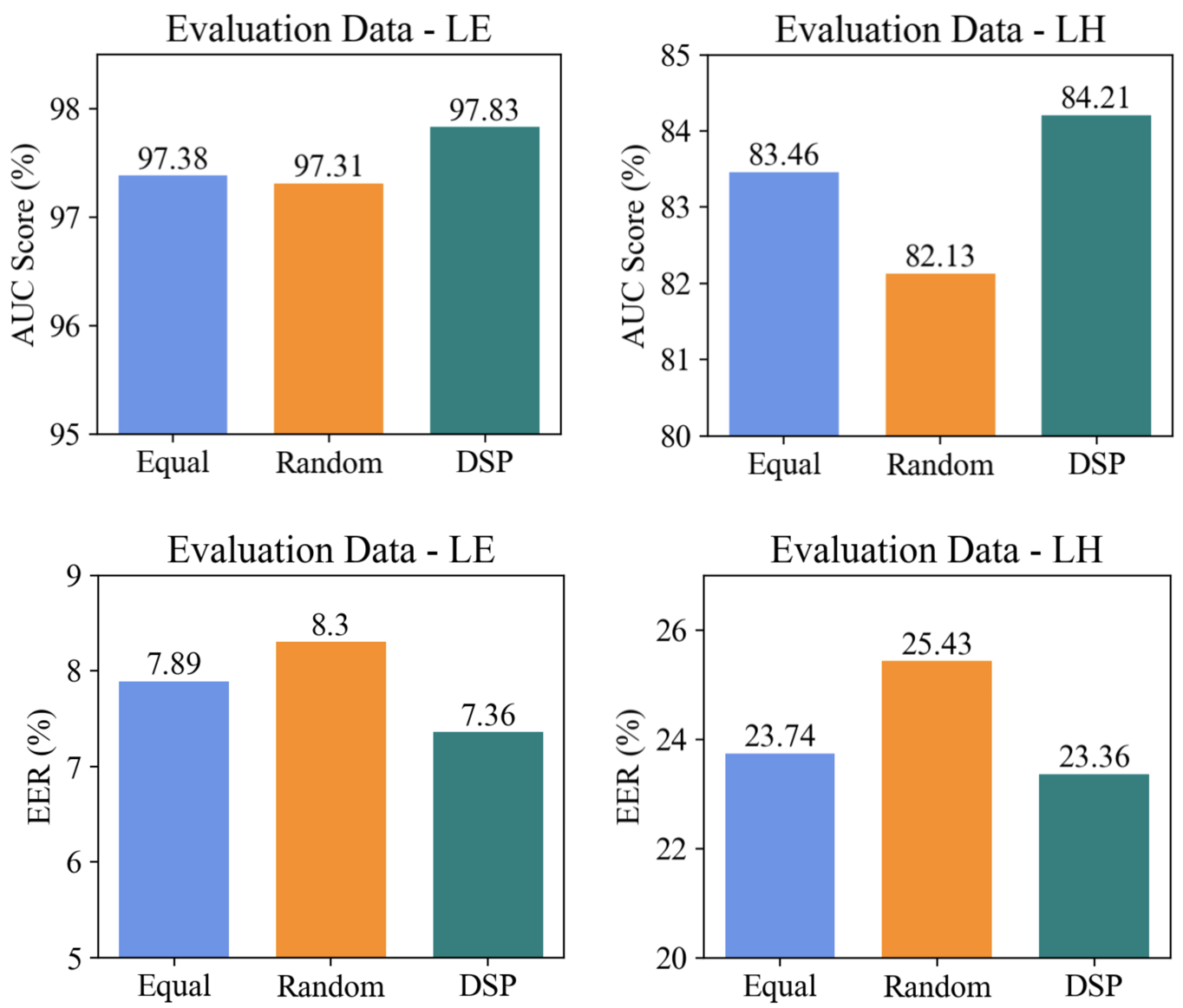}
      \caption{Experimental results using different sequence partitioning schemes employed to partition the audio sequence to match its length with the text sequence. Proposed DSP algorithm shows superior performance on both LE and LH datasets in terms of AUC Score and EER metric.}
      \label{fig:partition}
    \end{figure}

    \begin{figure}[t!]
      \centering
     \begin{subfigure}[b]{0.5\linewidth}
         \centering
         \includegraphics[width=\linewidth]{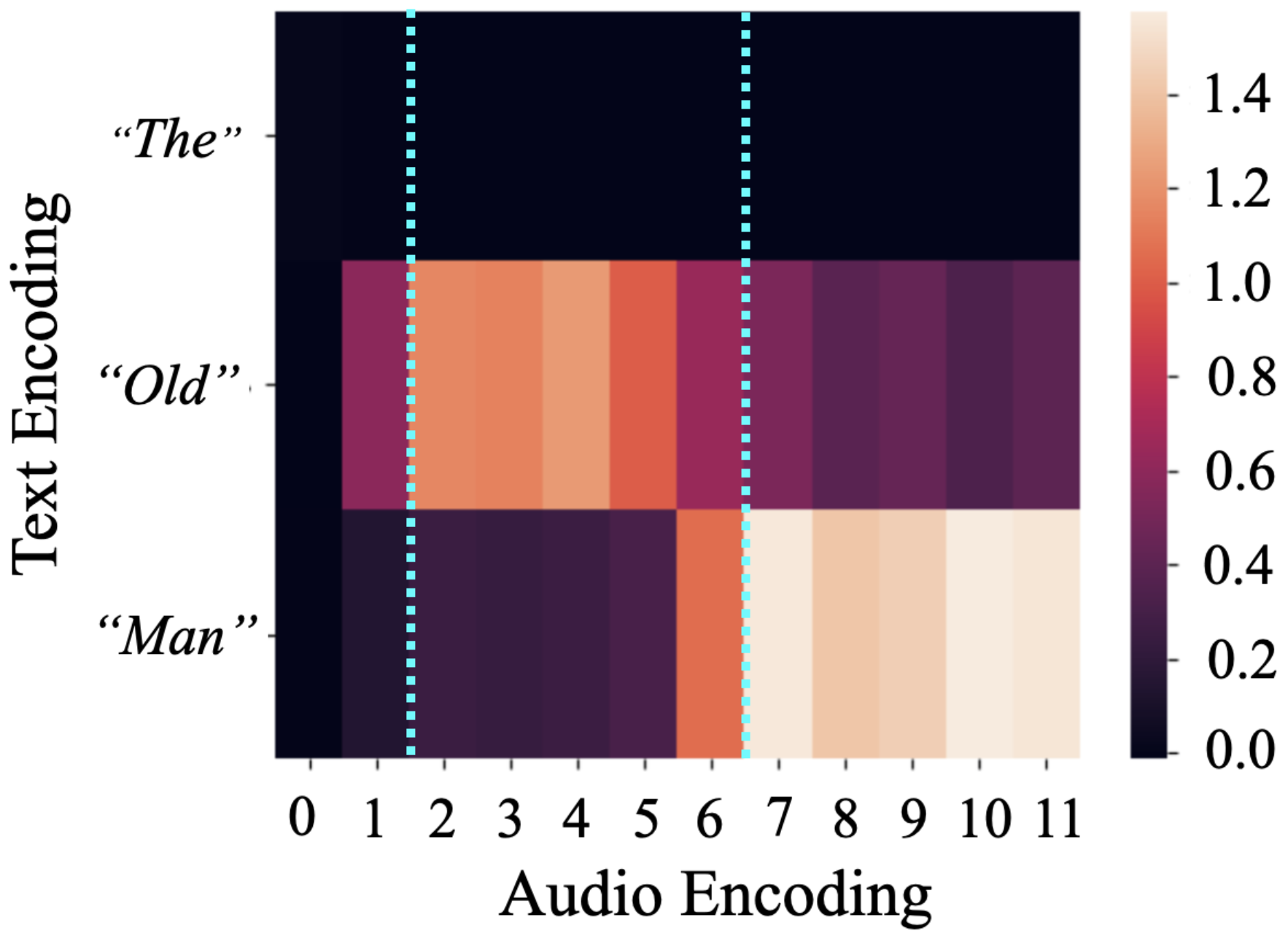}
         \caption{}
         \label{fig:3worda}
     \end{subfigure}
     \begin{subfigure}[b]{0.45\linewidth}
         \centering
         \includegraphics[width=\linewidth]{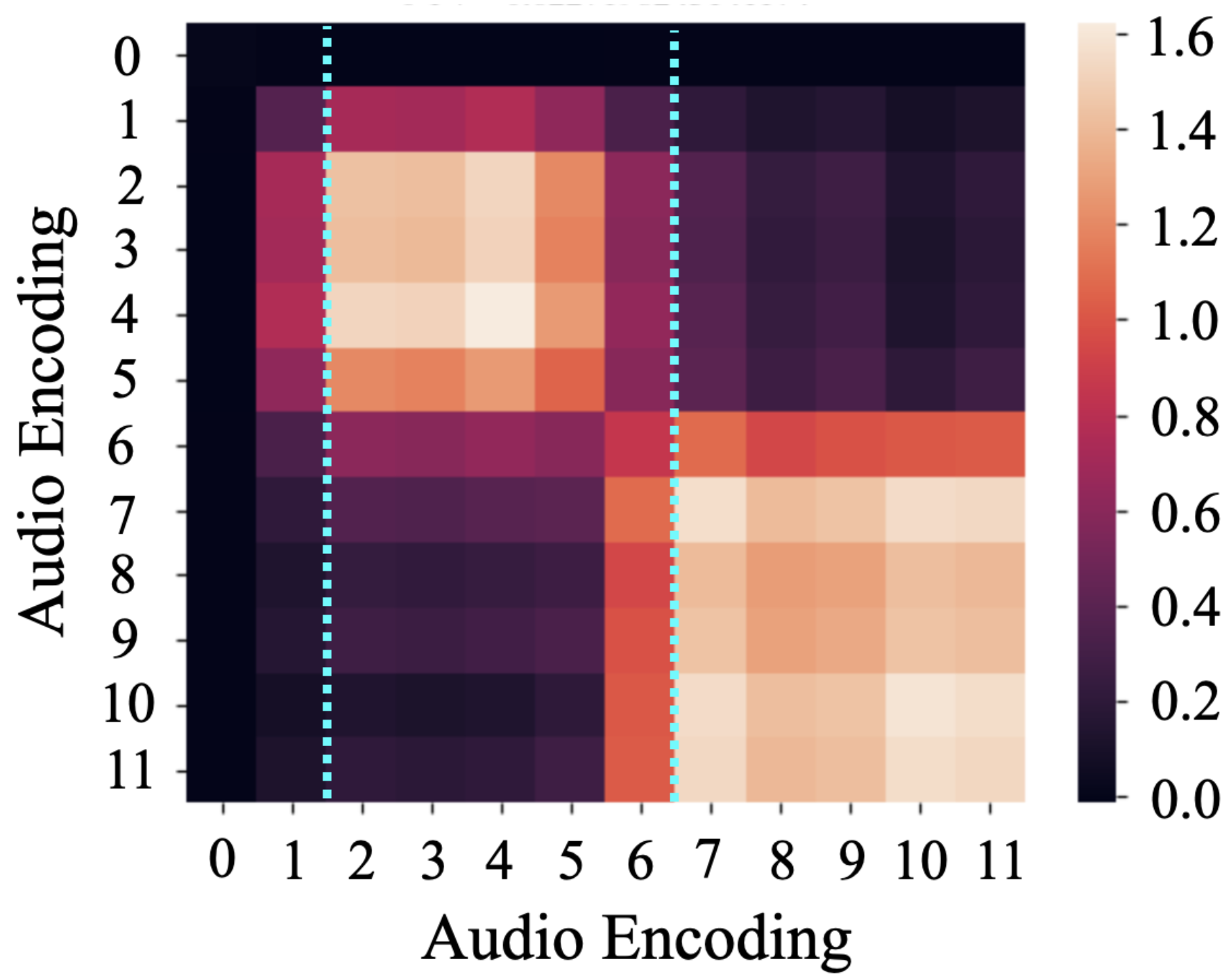}
         \caption{}
         \label{fig:3wordb}
     \end{subfigure}
     \begin{subfigure}[b]{0.48\linewidth}
         \centering
         \includegraphics[width=\linewidth]{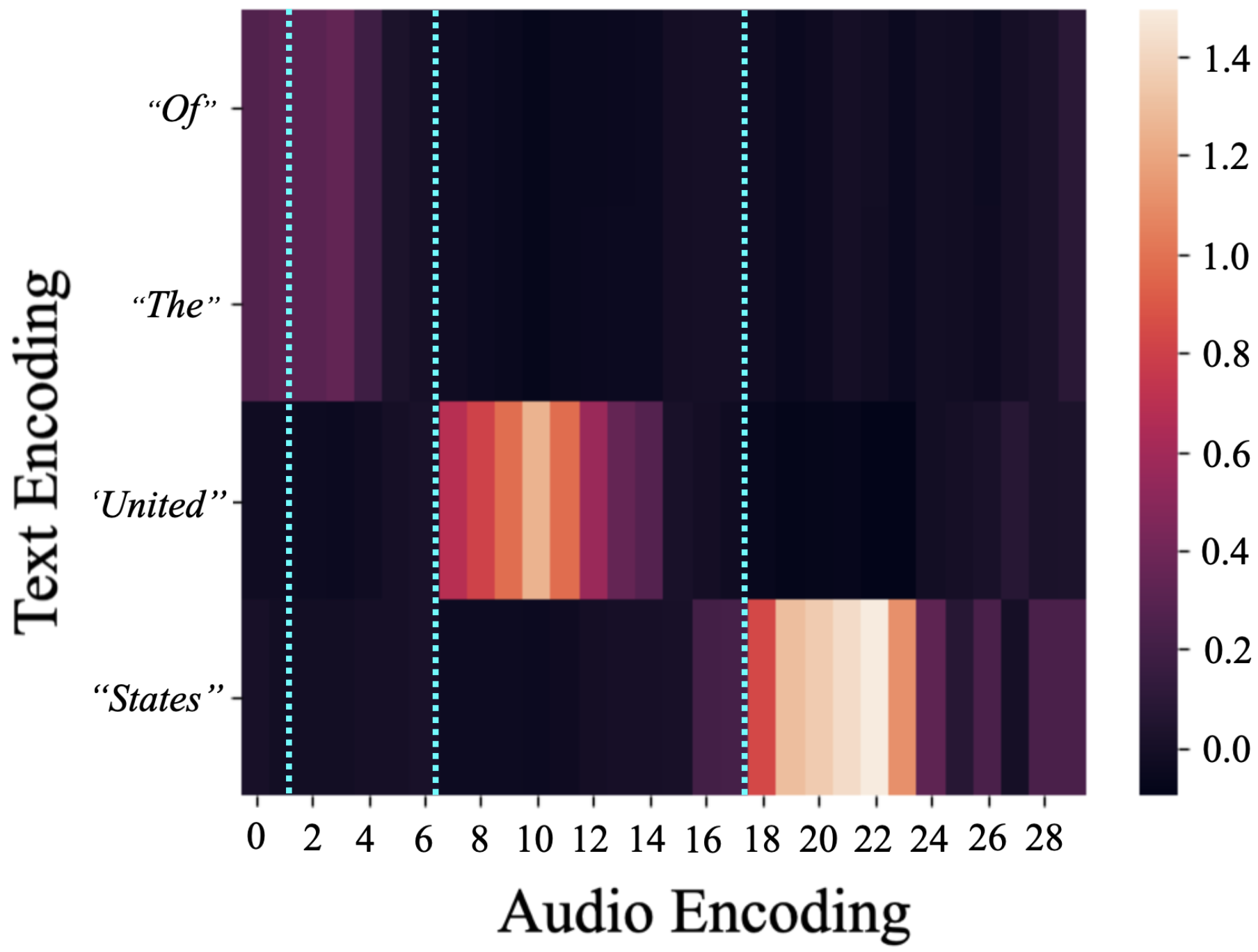}
         \caption{}
         \label{fig:4worda}
     \end{subfigure}
     \begin{subfigure}[b]{0.45\linewidth}
         \centering
         \includegraphics[width=\linewidth]{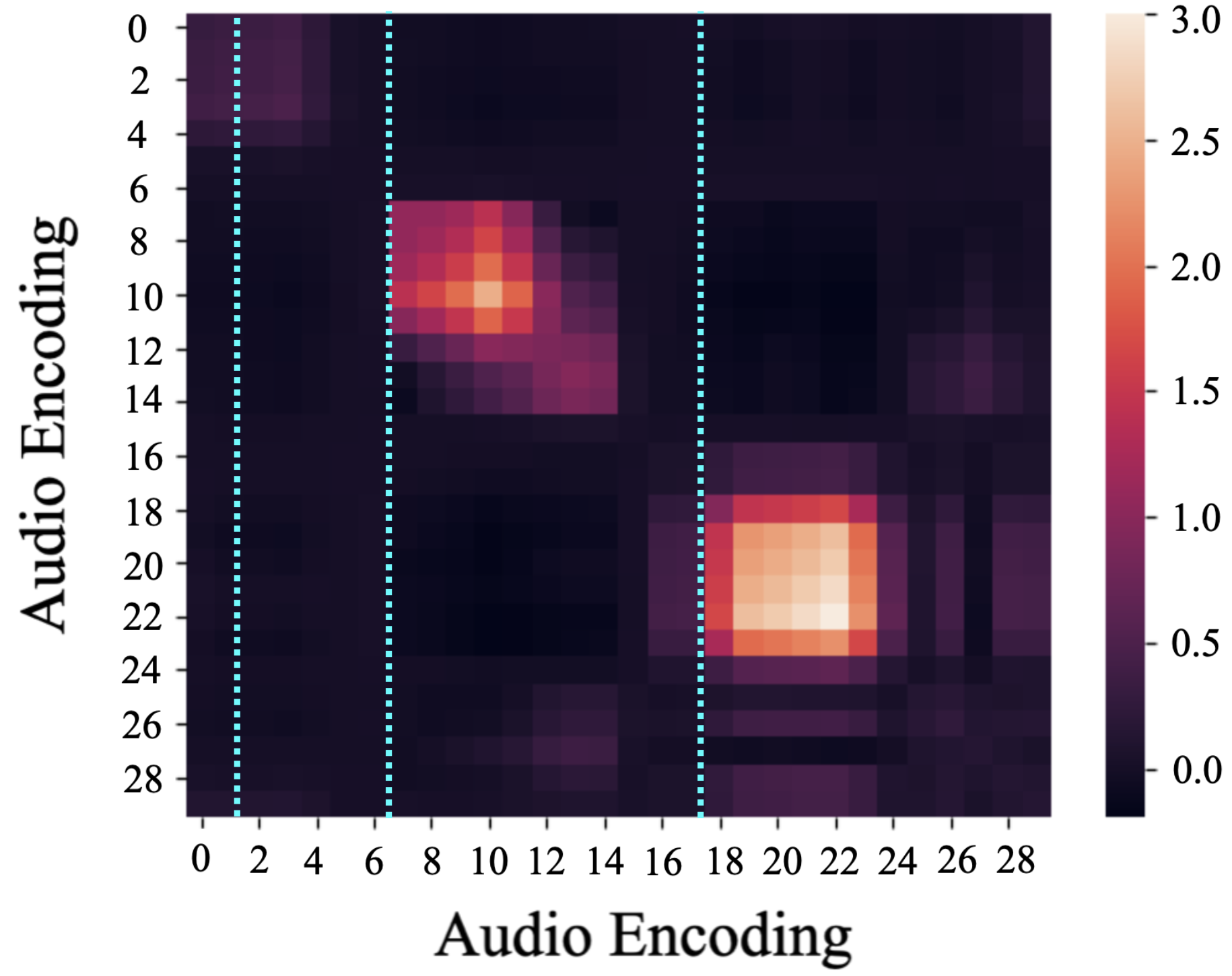}
         \caption{}
         \label{fig:4wordb}
     \end{subfigure}
     \caption{Visualization of the projected audio and text embeddings using correlation between text-to-audio and audio-to-audio embeddings. (a) and (b) are for a positive example, ``The Old Man". (c) and (d) are for a positive example, ``Of The United States"}
     \label{fig:aligner}     
    \end{figure}


\section{Conclusions}
In this paper, we proposed an audio-text-based end-to-end model architecture for flexible keyword spotting, consisting of encoder, projector, and audio-text aligner modules. 
We address the key challenge of comparing two sequences of different lengths by proposing a novel dynamic programming-based algorithm, Dynamic Sequence Partitioning (DSP), to optimally partition the audio embedding sequence into the same length as the text sequence.
Experimental results showed that the proposed architecture outperformed the state-of-the-art results.


\bibliographystyle{IEEEtran}
\bibliography{paper}

\end{document}